\documentclass[preprint]{elsarticle}

\usepackage{hyperref}
\usepackage{geometry}
\biboptions{authoryear}
\usepackage{multicol}

\usepackage{csquotes}
\usepackage{amsmath}
\usepackage{amssymb}
\usepackage[official]{eurosym}
\usepackage{bm}
\usepackage{bbm}
\usepackage{url}

\usepackage{graphicx}
\graphicspath{{}}

\makeatletter
\def\ps@pprintTitle{%
 \let\@oddhead\@empty
 \let\@evenhead\@empty
 \def\@oddfoot{\centerline{Submitted Manuscript}}%
 \let\@evenfoot\@oddfoot}
\makeatother

\begin{document}

\begin{frontmatter}
\title{Importance Subsampling: Improving Power System Planning Under Climate-based Uncertainty}
\author[imperial]{Adriaan P Hilbers\corref{mycorrespondingauthor}}
\cortext[mycorrespondingauthor]{Corresponding author}
\ead{aph416@ic.ac.uk}
\author[reading,centre]{David J Brayshaw}
\author[imperial]{Axel Gandy}
\address[imperial]{Department of Mathematics, Imperial College London}
\address[reading]{Department of Meteorology, University of Reading}
\address[centre]{National Centre for Atmospheric Science, University of Reading}

\begin{abstract}
\includegraphics[scale=0.69, trim=0 0 0 0, clip]{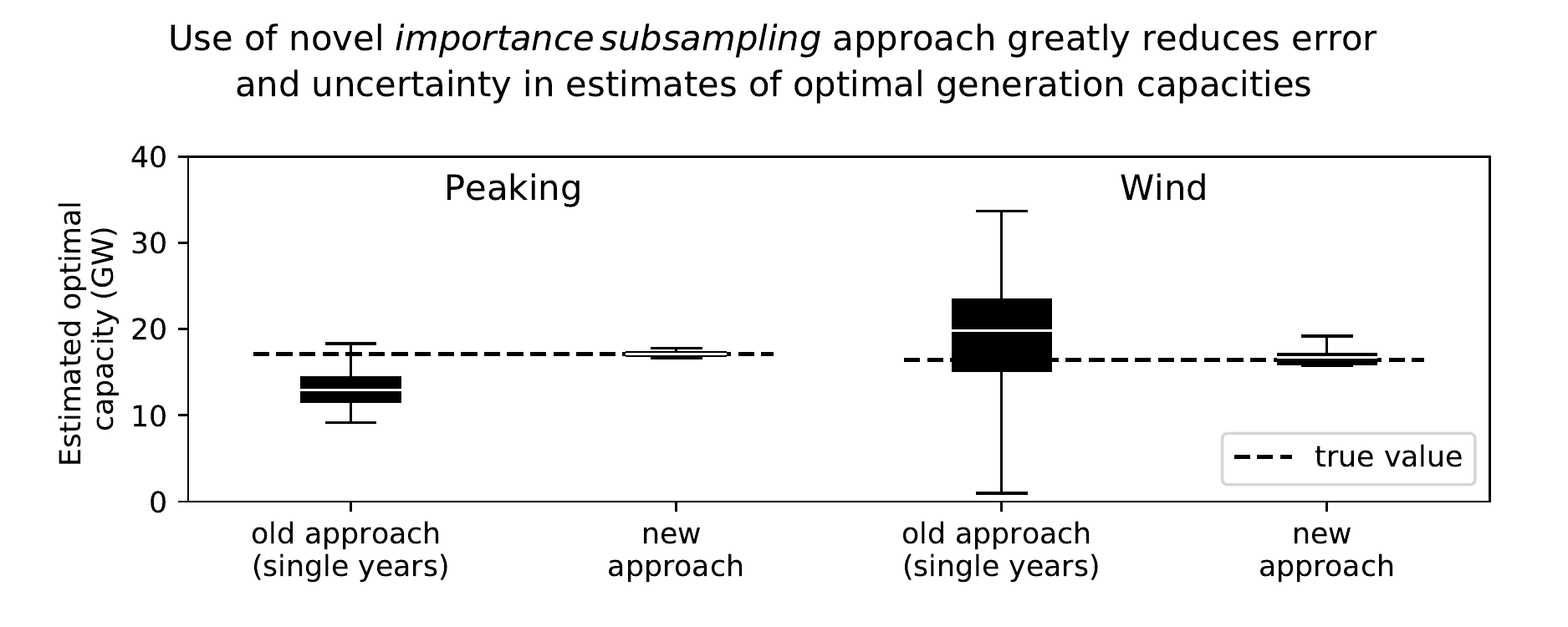}
  
Recent studies indicate that the effects of inter-annual climate-based variability in power system planning are significant and that long samples of demand \& weather data (spanning multiple decades) should be considered. At the same time, modelling renewable generation such as solar and wind requires high temporal resolution to capture fluctuations in output levels. In many realistic power system models, using long samples at high temporal resolution is computationally unfeasible. This paper introduces a novel subsampling approach, referred to as \textit{importance subsampling}, allowing the use of multiple decades of demand \& weather data in power system planning models at reduced computational cost. The methodology can be applied in a wide class of optimisation-based power system simulations. A test case is performed on a model of the United Kingdom created using the open-source modelling framework \textit{Calliope} and 36 years of hourly demand and wind data. Standard data reduction approaches such as using individual years or clustering into representative days lead to significant errors in estimates of optimal system design. Furthermore, the resultant power systems lead to supply capacity shortages, raising questions of generation capacity adequacy. In contrast, \textit{importance subsampling} leads to accurate estimates of optimal system design at greatly reduced computational cost, with resultant power systems able to meet demand across all 36 years of demand \& weather scenarios.
\end{abstract}


\end{frontmatter}

\section{Introduction}
\label{sec:introduction}
\subsection{Context}
\label{sec:introduction:context}
In the face of climate change, worldwide efforts are being undertaken to reduce carbon emissions \citep{ipcc_2018}. A common roadmap to sustainability is to decarbonise electricity supply and electrify other sectors such as as transport and heating \citep{staffel_2017}. In many countries, an essential part of this strategy is an increased use of variable renewable energy (VRE) generation such as solar and wind \citep{jacobson_2015, fes_2017}. Hence, while electricity demand has always exhibited some weather-dependence \citep[see e.g.][]{thornton_2017}, the same increasingly holds for supply.

To aid in questions of energy strategy, decision-makers frequently employ energy system models (ESMs), computer programs that simulate the energy transitions in a given geographical region \citep{ventosa_2005}. Power system models (PSMs) form a subset concerned primarily with the electricity sector. A common use of PSMs is to calculate the optimal \textit{generation mix} by minimising the sum of installation and generation costs while meeting demand \citep{stoft_book, bazmi_2011, ringkjob_2018}. Models considering renewables require coherent weather timeseries such as windspeeds or solar irradiances as inputs.

\subsection{The computational cost of weather \& climate variability}
\label{sec:introduction:computational_cost}
Recent studies indicate that robust power system planning under natural climate variability requires long samples of demand and weather data (spanning multiple decades). In particular, characteristics of power systems, such as the optimal installed capacities of different generation technologies, may be highly dependent on which year of data is used. Hence, power systems designed using single-year simulations may be suboptimal in the long run \citep{bloomfield_2016, pfenninger_2017, zeyringer_2018, wohland_2018}. The degree of this dependence is expected to increase as more VRE generation is employed \citep{staffel_2018, collins_2018}.

The need for long samples poses a computational challenge since accurately modelling systems with significant shares of VRE generation also requires high temporal resolution. Previous studies indicate that models with low resolution fail to capture VRE output fluctuations and underestimate required flexible and dispatchable generation capacity \citep{de_jonghe_2011, poncelet_2016, collins_2017, kotzur_2018}. For many realistic PSMs, it is computationally unfeasible to solve the optimisation problem using both a long sample of data and a high temporal resolution \citep{pfenninger_2017}.

\subsection{Established data reduction approaches}
\label{sec:introduction:data_reduction}
Various approaches to reduce computational cost without lowering temporal resolution exist. One strategy is the \textit{soft-linking} of a long-term planning model with a more detailed simulation operating on a shorter scale \citep{ringkjob_2018, zeyringer_2018, collins_2017}. Another is to run a model with a smaller number of representative periods (e.g. days or weeks) obtained by clustering the full dataset. Numerous studies explore the efficacy of such approaches in reproducing model outputs at reduced computational expense \citep{sisternes_2013, pfenninger_2017, nahmmaccher_2016, kotzur_2018, hartel_2017, poncelet_2017}. They arrive at a number of common conclusions. Firstly, the reduction approach must not remove extremes (e.g. by ``averaging away'' peaks) since including them ensures the model determines power system design able to meet demand in such scenarios. For this reason, heuristic adjustments such as including the maximum demand day are sometimes employed. Furthermore, clustering typically works poorly when applied to multiple decades of timeseries data since small changes in approach may lead to large spreads in model outputs. For example, \citet{pfenninger_2017} clusters 25 years of demand \& weather data and finds that optimal wind capacity is anywhere between 0.8 and 2.8 times peak demand depending on the choice of clustering algorithm and heuristic adjustment, giving the user virtually no indication of a good investment strategy. In addition, clustering does not generalise easily to models taking a large number of input timeseries. Consider a model with hourly resolution for 10 demand regions, 5 wind farms and 5 concentrated solar power plants. Clustering the days requires clustering vectors of length 480 (=24$\cdot$(10+5+5)).

\subsection{This paper's contribution}
\label{sec:introduction:contribution}
This paper introduces a novel subsampling approach, called \textit{importance subsampling}, that can be applied to multiple decades of demand \& weather timeseries data. PSM outputs evaluated using subsamples reliably estimate those found using the full timeseries at greatly reduced computational cost. The methodology is introduced in full generality and can be either directly applied or straightforwardly generalised to a wide class of optimisation-based PSMs.

A test case is performed using a model of the United Kingdom power system created using the open-source energy modelling framework \textit{Calliope} \citep[see][]{pfenninger_2018} and 36 years of UK-wide demand and wind data. Model outputs using \textit{importance subsampling} reproduce those using all 36 years of data more reliably than alternative approaches such as using individual years or clustering days.

This paper is structured as follows. Section \ref{doc_motivation} provides motivation, outlining the risks in designing power systems using short samples of demand \& weather data. Section \ref{doc_method} describes the \textit{importance subsampling} approach in detail and full generality. Section \ref{doc_tc} evaluates its use in a test case, comparing results to those found using individual years or by clustering timeseries into a number of representative days. Section \ref{doc_discussion} discusses the results' implications and recommends potential extensions. In the appendix (Section \ref{doc_app}), the reader finds full descriptions of the PSM and datasets employed.

\section{Motivation: climate variability in power system planning}
\label{doc_motivation}

\begin{figure}
  \hspace{2em} (a) Optimal generation capacity \hspace{3.5em} (b) Annual hours of unmet demand \\
  \vspace{0em} \\
\begin{tabular}{c}
  \includegraphics[scale=0.7, trim=10 0 10 10, clip]{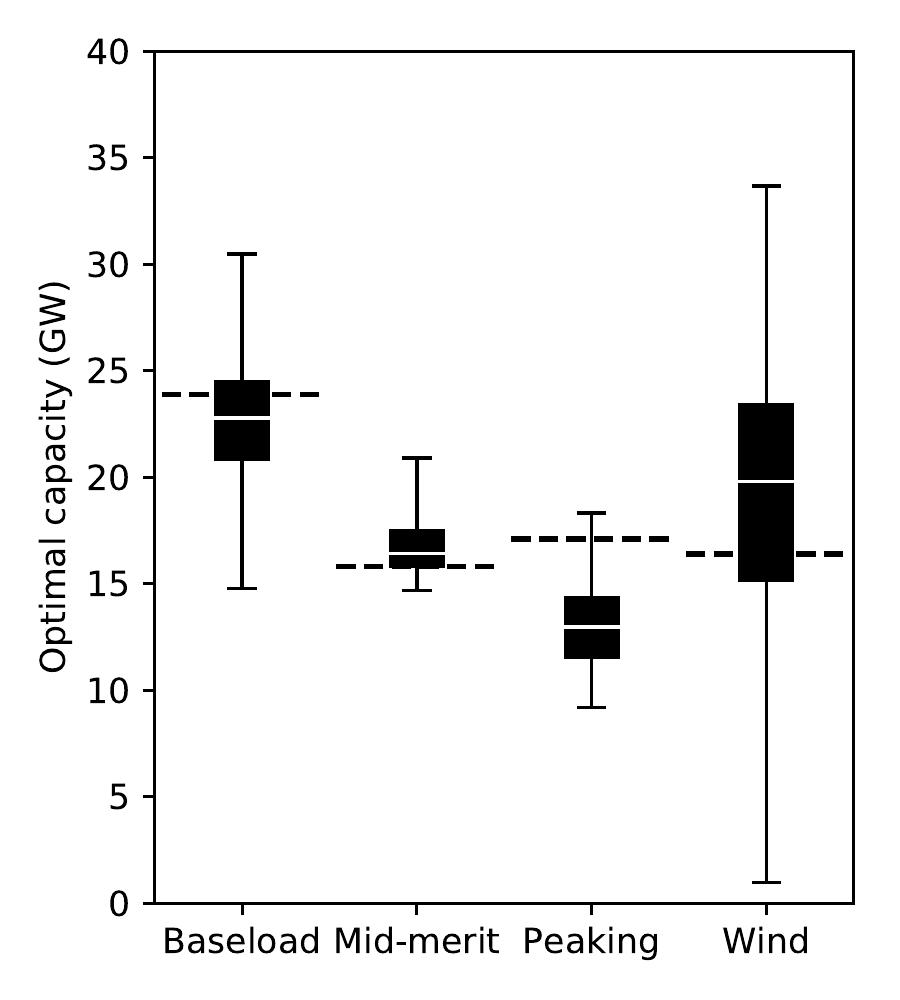}
\end{tabular} \hspace{0.5em}
\begin{tabular}{c}
  \includegraphics[scale=0.7, trim=10 0 0 23, clip]{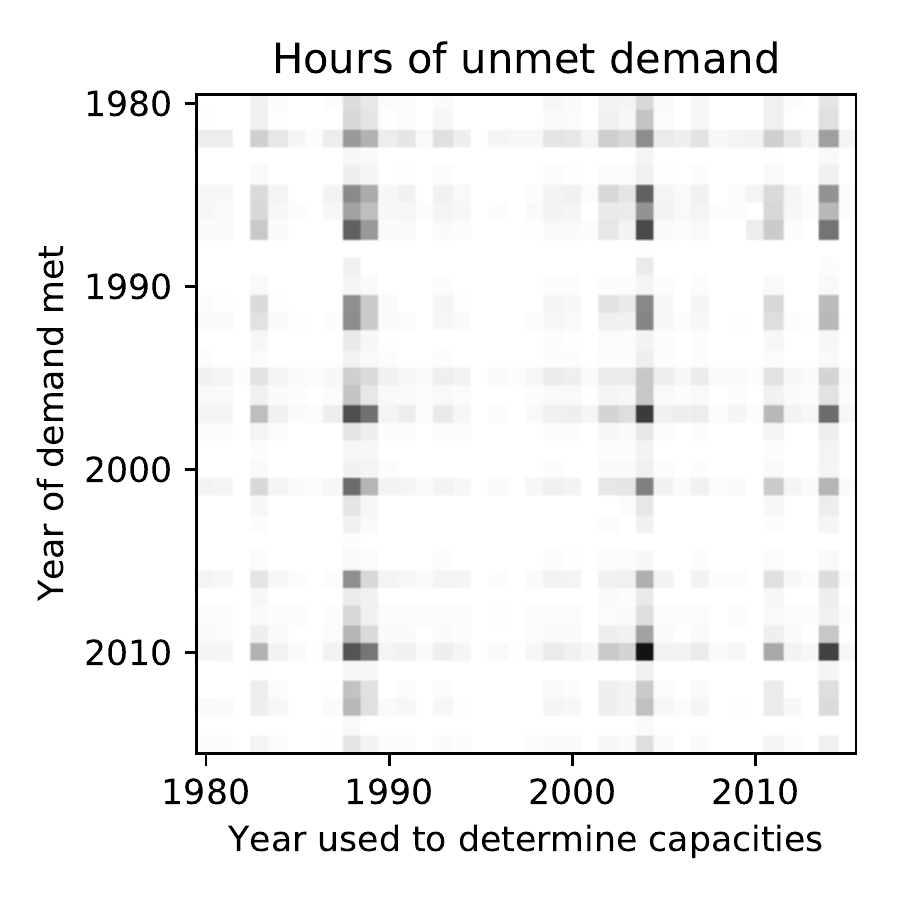} \\ \hspace{2.5em} \includegraphics[scale=0.66, trim=0 0 0 0, clip]{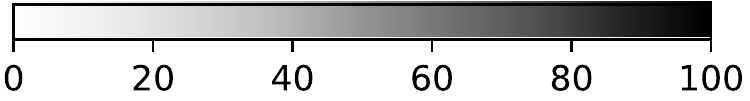}
\end{tabular}
\caption{(a) Distribution of optimal capacities across individual-year simulations from 1980 to 2015. The box shows the 25th, median and 75th percentile, and the whiskers show the minimum and maximum. The optimal capacities for a simulation using all 36 years of data are shown as a wider dashed line for comparison and serve as the best guess for the ``true'' optima. (b) Number of hours with supply capacity shortages. The $x$-axis indicates the year of data used to determine the optimal capacities and the $y$-axis the year of demand \& weather data in which to  meet demand. For example, the top right square shows the number of hours in which a system designed using 2015's optimal capacities fails to meet demand when used in the year 1980.}
\label{fig:motivation}
\end{figure}

This section provides a concrete illustration of the problem of inter-year climate-based variability in power system planning. A PSM of the UK is considered along with 36 years of hourly demand and wind levels (over the period 1980 to 2015). The demand timeseries is detrended so that the only differences between years are climate-driven (e.g. demand being higher when it is cold, see Section \ref{doc_app_data_demand} for details).

The box-and-whisker plot in figure \ref{fig:motivation}(a) shows the distribution of optimal generation capacities of 4 technology types across the 36 individual years of historic data (i.e. for 1980, 1981, ... , 2014, 2015). The white horizontal line inside the box indicates the median and the wider dashed horizontal line shows the optimal capacity when all 36 years of data considered at once: the best estimate for the ``truly'' optimal design. Two phenomena stand out. The first is the considerable inter-year spread, particularly for wind. Using 2010 data indicates almost no wind at all should be built (1.0GW), whereas the optimal configuration for 1986 has more wind capacity than any other technology (33.7GW). The second is the bias between the median of individual-year model runs and the 36-year optimal capacities (seen by comparing the white line on the box plots with the wider dashed black line). For example, 92\% of individual-year model runs underestimate optimal peaking capacity, with a median underestimation of around 20\%. The opposite is true for wind, for which the median individual-year wind capacity overestimates the 36-year optimum by around 15\%. 

Furthermore, a power system designed using one year's optimal capacities may lead to supply capacity shortages in another. Figure \ref{fig:motivation}(b) shows the distribution of the number of hours with supply capacity shortages (insufficient capacity to meet demand) as a function of the year used to determine the optimal capacities (in other words, the system is ``designed'' on basis of the data of the year indicated on the $x$-axis and asked to meet conditions of the year on the $y$-axis). The risk of picking the ``wrong year'' on which to base system design is high: one designed using 2004's capacities fails to meet 93 hours of demand in 2010, and an average of 30 hours annually across all 36 years. Across the whole dataset, a power system designed using a random choice of year has a 50\% chance of at least one hour of supply capacity shortage in another, and a 35\% chance of shortage for at least 3 hours. This highlights the risks of informing power system strategy on single years and illustrates the need to consider longer samples.

\section{Methodology}
\label{doc_method}
This section introduces the precise \textit{importance subsampling} methodology in full generality. Like the clustering strategies discussed in Section \ref{sec:introduction:data_reduction}, it is based on timeseries compression through subsampling. The specific application to a power system planning problem is presented in Section \ref{doc_tc}.

\subsection{Intuition}
\label{sec:method:intuition}

The \textit{importance subsampling} methodology works by subsampling timesteps from the full dataset. They are always selected in full; if one timeseries value at timestep $t$ is sampled, then so are all others. This ensures correlations between timeseries (e.g. spatial windspeed correlations or demand-wind correlations) are automatically accounted for.

Determining which timesteps to sample can be complicated. In Section \ref{doc_motivation}, a system designed using 2004 data leads to supply capacity shortages in other years because all of 2004's high-demand timesteps have high wind levels. As a result, the 2004 optimal design underestimates the amount of required backup (dispatchable) generation capacity. This highlights the fact that cost-optimal power system design will never have excess capacity; it will \textit{just} be able to meet demand in the ``worst'' timestep (the one requiring the largest supply). Indeed, this is the reason that heuristic adjustments such as including the maximum demand day are sometimes employed in the established subsampling approaches outlined in Section \ref{sec:introduction:data_reduction}. The disadvantage of heuristic methods is that, when VRE generation is involved, they may fail to identify those timesteps truly required to ensure generation capacity adequacy. The peak demand in 2004 is not lower than in other years, so including the maximum demand day fails to mitigate the underestimation of required generation capacity. In this case, it may be necessary to sample instead a day with a slightly lower demand but virtually no wind. Determining the correct trade-off between high demand and low wind (e.g. is a timestep with 50GW demand and 0.1 wind capacity factor ``worse'' than one with 55GW peak demand and 0.2 wind capacity factor?) is difficult \textit{a priori} and will differ between PSMs.

The \textit{importance subsampling} approach identifies the essential timesteps in a systematic way by assigning each one a measure of the difficulty in meeting demand, referred to as its \textit{importance}. This has two advantages. Firstly, it one-dimensionalises timeseries inputs: a timestep with many values (e.g. demand levels and renewable outputs across multiple regions) is assigned a single \textit{importance}. This allows timesteps to be ranked objectively irrespective of the PSM's structure. Secondly, the ranking means a selection of timestep(s) with the highest \textit{importance} can be included forcibly into the modelling sample to ensure generation capacity adequacy in the resultant power system.

A timestep's \textit{importance} may depend on the power system design. In single-region models, the \textit{net demand} (residual demand after all renewable generation is used) is a good candidate since it is equal to the required dispatchable generation. The net demand, however, depends on the installed renewable capacity. For example, a timestep with a demand of 50GW and wind capacity factor 0.1 has a higher net demand than one with a demand of 55GW and wind capacity factor of 0.3 exactly when the installed wind capacity is less than 25GW. Since the installed wind capacity is itself a model output, a \enquote{Catch 22} situation occurs: determining which timesteps to use in a simulation requires an estimate for each timestep's \textit{importance}, which itself requires the model outputs.

A two-stage approach is therefore proposed. A stage 1 optimisation run, using a random sample of timesteps, gives a rough indication of optimal power system design. This is used to estimate each timestep's \textit{importance}. A stage 2 sample is subsequently created by including a number of the timesteps with the highest \textit{importance} and a random selection of those remaining. This is used in a second model run to to estimate the model outputs found using the full dataset. The approach is dependent on the choice of \textit{importance} function, which should be a proxy for the ``difficulty'' in meeting demand (in terms of requiring a large amount of generation capacity). In the case study (Section \ref{doc_tc}), a timestep's \textit{variable cost} is used and this choice is applicable in a large class of PSMs. However, others are possible, and the choice may be tailored through consideration of the PSM or from expert knowledge.

\subsection{Precise methodology}
\label{sec:method:precise}

\begin{figure}
\centering
\footnotesize
\begin{tabular}{c}
Full timeseries data: $(\text{ts\_data}_t)_{t \in \mathcal{T}^F}$ \\
\tiny{$\downarrow$} \\
Random subsample: $(\text{ts\_data}_t)_{t \in \mathcal{T}^{S1} \subset \mathcal{T}^F}$ \\
\tiny{$\downarrow$} \\
Stage 1 estimate of optimal design: \\
$\text{design}_\text{opt}^{S1} = \text{PSM}((\text{ts\_data}_t, \text{weight}_t)_{t \in \mathcal{T}^{S1}})$ \\
\tiny{$\downarrow$} \\
\textit{Importance} of each timestep: \\
$\text{imp}_t = \text{IMP}(\text{ts\_data}_t, \text{design}_\text{opt}^{S1}) \hspace{0.5em} \forall t \in \mathcal{T}^F$ \\
\tiny{$\downarrow$} \\
\textit{Importance} subsample (see figure): \\
$(\text{ts\_data}_t, \text{weight}_t)_{t \in \mathcal{T}^{S2} \subset \mathcal{T}^F}$ \\
\tiny{$\downarrow$} \\
Stage 2 estimate of optimal design: \\
$\text{design}_\text{opt}^{S2} = \text{PSM}((\text{ts\_data}_t, \text{weight}_t)_{t \in \mathcal{T}^{S2}})$ \\
\end{tabular}
\begin{tabular}{c}
\includegraphics[scale=0.72, trim=12 10 10 0, clip]{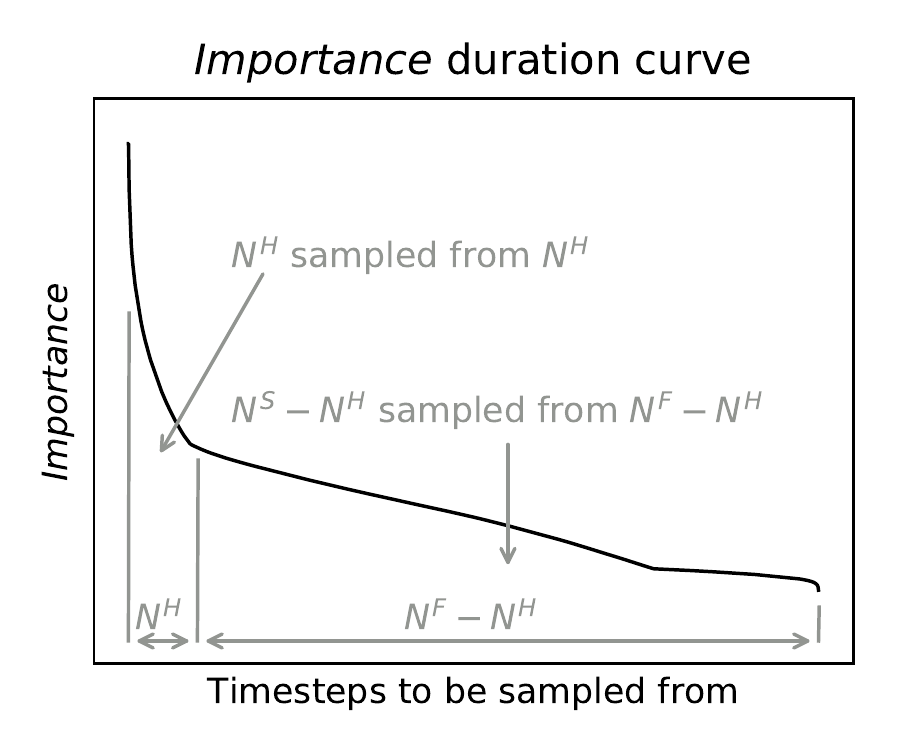}
\end{tabular}
\caption{\textit{Importance subsampling} methodology. The \textit{importance} duration curve is constructed by plotting timesteps in descending order of \textit{importance} in direct analogy to the \textit{load duration curve} common across power system analysis. The \textit{importance subsample} contains, from the full timeseries, the $N^H$ timesteps with the highest \textit{importance} and a random selection of size $N^S - N^H$ from those remaining. Timesteps are weighted to reflect their relative proportions across the full dataset as described in Section \ref{sec:method:precise}.}
\label{fig:method_illustration}
\end{figure}

The following terminology is employed throughout this section:
\begin{itemize}
\item $\mathcal{T}^F = \{1, \ldots, N^F\}$: full set of timesteps to be sampled from.
\item $\text{ts\_data}_t$: timeseries input data in timestep $t$. For example, in a model with demand levels and wind capacity factors, $\text{ts\_data}_t = (\text{demand}_t, \text{wind}_t)$.
\item $\text{weight}_t$: weight assigned to timestep $t$ in a PSM. For example, when clustering timeseries into representative periods (see Section \ref{sec:introduction:data_reduction}), periods are typically weighted for their relative cluster sizes.
\item $\text{design}_\text{opt}$: optimal power system design, e.g. a vector with each generation technology's installed capacity.
\end{itemize}
One furthermore needs two functions:
\begin{itemize}
\item A power system model with an optimiser that returns, given some demand \& weather timeseries data and timestep weights indexed by timesteps $\mathcal{T}$, the optimal power system design:
\begin{equation}
\text{design}_\text{opt} = \text{PSM} ((\text{ts\_data}_t, \text{weight}_t)_{t \in \mathcal{T}}).
\end{equation}
\item An \textit{importance} function that assigns to each timestep a measure of the difficulty in meeting demand. This may depend on the power system design:
\begin{equation}
\text{imp}_t = \text{IMP}(\text{ts\_data}_t, \text{design}_\text{opt}).
\end{equation}
\end{itemize}

\noindent \textit{Importance subsampling} works as follows. Suppose one wants to determine, given some (equally weighted) timeseries data $(\text{ts\_data}_t, \text{weight}_t)_{t \in \mathcal{T}^F}$, the optimal power system design $\text{design}_\text{opt}^F = \text{PSM}((\text{ts\_data}_t, \text{weight}_t)_{t \in \mathcal{T}^F})$. Evaluating this function may be too computationally expensive, so $\text{design}_\text{opt}^F$ is estimated as follows:
\begin{enumerate}
\item Randomly sample $N^S$ timesteps from the $N^F$ to create a stage 1 subsample $\mathcal{T^\text{S1}} \subset \mathcal{T}^F$ of length $N^S$ with equal weights.
\item Determine optimal design for stage 1 subsample:
\begin{equation}
\text{design}_\text{opt}^\text{S1} = \text{PSM}((\text{ts\_data}_t, \text{weight}_t)_{t \in \mathcal{T}^\text{S1}}).
\end{equation}
\item Calculate \textit{importance} of each timestep in the full dataset using stage 1 design:
\begin{equation}
\text{imp}_t = \text{IMP}(\text{ts\_data}_t, \text{design}_\text{opt}^{S1}) \hspace{0.5em} \forall \, t \in \mathcal{T}^F.
\label{eq_method_cost}
\end{equation}
\item Create a stage 2 sample of size $N^S$ using the $N^H$ timesteps with the highest \textit{importance} and a random selection of those remaining:
\begin{equation}
\mathcal{T}^\text{S2} = \{\underbrace{t_1^\text{S2}, t_2^\text{S2}, \ldots, t_{N^H-1}^\text{S2},t_{N^H}^\text{S2}}_{\substack{\text{Timesteps with} \\ \text{highest \textit{importance}}}}, \underbrace{t_{N^H+1}^\text{S2}, t_{N^H+2}^\text{S2}, \ldots, t_{N^S-1}^\text{S2}, t_{N^S}^\text{S2}}_{\substack{\text{Random selection} \\ \text{of those remaining}}}\} \subset \mathcal{T}^F
\end{equation}
with associated weights
\begin{equation}
\text{weight}_t^\text{S2} = 
\begin{cases}
\frac{1}{N^F} \hspace{5.8em} t \in \{t_1^\text{S2}, \ldots, t_{N^H}^\text{S2}\} \\
\frac{N^F - N^H}{N^F(N^S - N^H)} \hspace{2em} t \in \{t_{N^H+1}^\text{S2}, \ldots, t_{N^S}^\text{S2}\}.
\end{cases}
\end{equation}
Figure \ref{fig:method_illustration} provides an illustration of this step.
\item Determine optimal design for stage 2 subsample:
\begin{equation}
  \text{design}_\text{opt}^{S2} = \text{PSM}((\text{ts\_data}_t, \text{weight}_t)_{t \in \mathcal{T}^{S2}
}).
\end{equation}
$\text{design}_\text{opt}^\text{S2}$ is the estimate of $\text{design}_\text{opt}^F$. 
\end{enumerate}

Each evaluation of $\text{design}_\text{opt}^\text{S2}$ involves 2 evaluations of optimal design (stage 1 and stage 2) using $N^S$ timesteps each. Since the computational cost of the other steps (sampling and evaluating \textit{importance}) is typically negligible, the cost of evaluating $\text{design}_\text{opt}^{S2}$ is twice that of a single evaluation with sample size $N^S$.  

The weights in step 4 account for the relative proportions of the two bins. For example, suppose an \textit{importance subsample} of size 120 timesteps is created using the 60 timesteps with the highest \textit{importance} and a random selection of 60 from those remaining. Half of this subsample consists of timesteps with a high \textit{importance} even though they represent a much smaller proportion of the full dataset. The weights are chosen to cancel out this oversampling and prevent overengineering for extreme scenarios. The $N^H$ timesteps with the highest \textit{importance} represent a proportion $\frac{N^H}{N^F}$ of the dataset and the other $N^S - N^H$ represent $\frac{N^F - N^H}{N^F}$. The weights account for this:
\begin{align}
\sum_{k=1}^{N^S} \text{weight}_{t_k^\text{S2}}^{S2} &= \sum_{k=1}^{N^H} \frac{1}{N^F} + \sum_{k=N^H + 1}^{N^S} \frac{N^F - N^H}{N^F(N^S - N^H)} \\
&= \underbrace{\frac{N^H}{N^F}}_{\substack{\text{Correct proportion} \\ \text{of total dataset}}} + \underbrace{\frac{N^F - N^H}{N^F}}_{\substack{\text{Correct proportion} \\ \text{of total dataset}}} = 1
\end{align}

\section{Test case: UK power system model}
\label{doc_tc}
\subsection{Overview}
\label{doc_tc_overview}

In this section, the performance of the \textit{importance subsampling} approach is compared with other subsampling strategies when applied to a PSM based on the United Kingdom (UK) and created in the open-source energy modelling framework \textit{Calliope} \citep[see][]{pfenninger_2018}. This model is designed as an idealised test case and should not be viewed as a realistic representation of the UK power system or used to inform policy or strategy. It is employed only since it uses a similar linear programming formulation as many PSMs popular in the energy community \citep[see e.g.][]{trutnevyte_2016, bazmi_2011, hall_2016} and results on this model can be reasonably expected to generalise to those models also.

Hourly UK-wide demand levels and wind capacity factors are estimated over the 36-year period from 1980 until 2015 and serve as the model's timeseries inputs. Long-term anthropogenic trends such as economic growth and efficiency improvements are removed so that different years of demand data can be fairly compared. Model outputs are the optimal installed capacities of 4 possible generation technologies. The first 3 (baseload, mid-merit and peaking) are generic dispatchable technologies that differ only in their installation and generation costs. The 4th technology, wind, has 0 generation cost but output capped by time-varying wind levels. The PSM and input timeseries are discussed fully in the appendix (Section \ref{doc_app}). 

A ``perfect model'' framework is assumed: the optimal capacities across the 36 years are taken to be those that minimise cost under the ``true'' distribution of demand and wind. The 36-year capacities (for baseload, mid-merit, peaking and wind technologies) hence serve as targets:
\begin{equation}
[\text{cap}_b^\text{36y}, \text{cap}_m^\text{36y}, \text{cap}_p^\text{36y}, \text{cap}_w^\text{36y}] = \text{PSM} ((d_t, w_t, \text{weight}_t)_{t \in \mathcal{T}^\text{36y}})
\label{eq_targets}
\end{equation}
where $d_t$ is the demand level and $w_t$ is the wind capacity factor in timestep $t$. Each timestep is assigned equal weight. A subsample $\hat{\mathcal{T}} \subset \mathcal{T}^\text{36y}$ generates estimators of the optimal capacities as defined by
\begin{equation}
[\hat{\text{cap}}_b, \hat{\text{cap}}_m, \hat{\text{cap}}_p, \hat{\text{cap}}_w] = \text{PSM} ((d_t, w_t, \text{weight}_t)_{t \in \hat{\mathcal{T}}}).
\label{eq_estimators}
\end{equation}
Subsampling strategies are evaluated on a number of criteria. The estimators defined by equation (\ref{eq_estimators}) should have both a low variation across samples generated by the same process and a low bias (understood as median error) when compared to the targets in equation (\ref{eq_targets}). Furthermore, suppose a power system is designed using the estimated optimal capacities in equation (\ref{eq_estimators}). For this (hypothetical) system, two statistics are calculated:
\begin{itemize}
\item \textit{hours of unmet demand}: number of hours in the full sample in which the power system has insufficient generation capacity to meet demand. For example, the optimal power system design for 1980 data may be unable to meet demand for some hours in the period 1981-2015, leading to hypothetical supply shortages.
\item \textit{extra system cost}: additional cost (sum of installation and generation costs) of meeting the 36-year demand using a suboptimal power system. A PSM user might be aware of the risks in designing systems using short samples and want to compensate. The cheapest way to do this \textit{a priori} (i.e. without more optimisation) is to use extra peaking capacity to meet any unmet demand. Define the \textit{extra system cost} of the sample $\hat{\mathcal{T}}$ by
\begin{equation}
\text{extra system cost}^{\hat{\mathcal{T}}} = (\text{system cost}^{\hat{\mathcal{T}}} - \text{system cost}^{\text{opt}})
\label{eq_extracost}
\end{equation}
where $\text{system cost}^{\hat{\mathcal{T}}}$ is the cost of a system designed using sample $\hat{\mathcal{T}}$ (with extra peaking capacity and generation to ensure no unmet demand) and $\text{system cost}^{\text{opt}}$ is the cost of the 36-year optimal system. Since it is by definition cost-optimal, equation (\ref{eq_extracost}) is nonnegative.
\end{itemize}

Computational cost increases with timeseries length, so samples should be made as short as possible. The computational expense in solving the optimisation problem in this model scales linearly in the number of timesteps. In some PSMs (e.g.\ mixed-integer linear programs), the computational effort increases faster than linearly, in which case a reduction in simulation length causes a proportionally larger decrease in computational cost.

Four subsampling strategies are investigated: using individual years, random sampling of timesteps, $k$-medoids clustering into representative days and \textit{importance subsampling}. In $k$-medoids clustering, samples are generated by clustering the 48-dimensional vectors of each individual day's (normalised) hourly demand and wind levels into $k$ clusters. Each cluster's representative day is the day in the full timeseries whose vector is closest to the cluster mean. The model is run on the $k$ representative days weighted by the number of days their cluster contains.

\subsection{\textit{Importance subsampling}: setup}
\label{doc_tc_sampling}
Recall from Section \ref{doc_method} that \textit{importance subsampling} requires a PSM that can determine optimal system design. The test case PSM is an example of this. Here, the design output is the vector of optimal installed capacities of the four technologies. It is determined by minimising the sum of installation and generation costs while meeting demand. Input data is hourly UK-wide demand and wind capacity factors:
\begin{align}
  \text{ts\_data}_t &= (d_t, w_t) \\
  \text{design}_\text{opt}
  &= [\text{cap}_b, \text{cap}_m, \text{cap}_p, \text{cap}_w]
\end{align}
where $d_t$ and $w_t$ are the demand level and wind capacity factor in timestep $t$ respectively. The second requirement is an \textit{importance function}. A timestep's \textit{variable cost} is proposed for this purpose:
\begin{align}
\text{imp}_t = \text{varcost}_t &= f_b \text{gen}_{bt} + f_m \text{gen}_{mt} + f_p \text{gen}_{pt} + f_w \text{gen}_{wt} \\
&=f_b \text{gen}_{bt} + f_m \text{gen}_{mt} + f_p \text{gen}_{pt}
\label{eq_varcost}
\end{align}
where $f_i$ is technology $i$'s cost per unit electricity generated (in \pounds/GWh) and $\text{gen}_{it}$ is the amount of electricity generated by technology $i$ in timestep $t$ (in GWh). The costs per unit electricity of the different technologies are shown in Table \ref{table_ra_costs} in the appendix. Wind is assumed to generate electricity at 0 marginal cost: $f_w=0$. 

\begin{figure}
\centering
\includegraphics[scale=0.7, trim=0 0 0 0, clip]{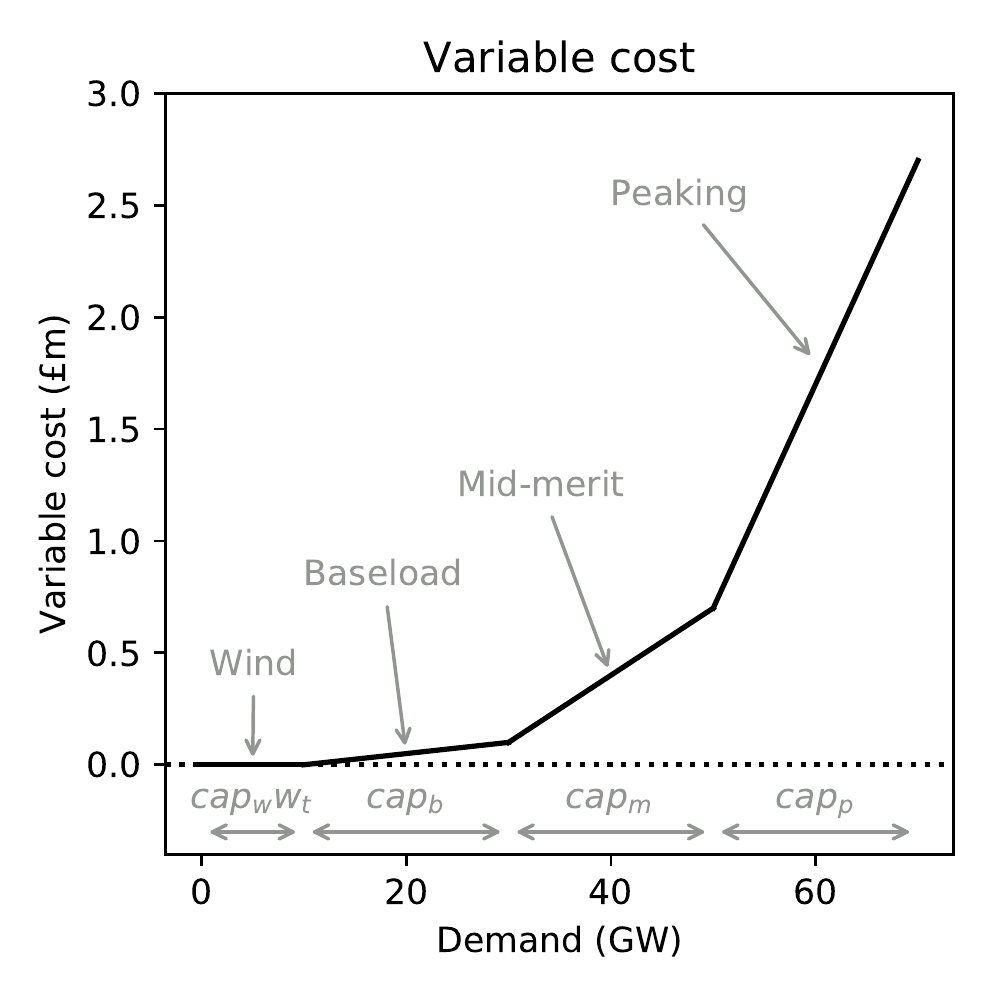}
\caption{A timesteps's variable cost as a function of demand. The first $\text{cap}_w w_t$ (installed wind capacity times wind capacity factor) GW are met by wind (with no contribution to variable cost), followed by baseload, mid-merit and peaking. The variable cost increases more quickly as progressively more expensive generation technologies are employed.}
\label{fig_sampling_varcost}
\end{figure}

A timestep's variable cost as a function of its demand is shown in Figure \ref{fig_sampling_varcost}. Demand is met by \textit{merit-order stacking} of technologies in ascending order of generation cost, with all available wind power used first, followed by baseload, mid-merit and peaking. The first $\text{cap}_w w_t$ GW (the installed capacity times the timestep's wind capacity factor) of demand are met by wind power with no contribution to the variable cost. The next $\text{cap}_b$ GW are met by baseload, followed by $\text{cap}_m$ GW of mid-merit and $\text{cap}_p$ GW of peaking (any unmet demand is also assumed to be met by peaking). The variable cost increases more quickly as technologies with progressively higher generation cost are used. It depends implicitly on the timestep's demand level and wind capacity factor, as well as the installed capacity of the different generation technologies, through the generation terms in equation (\ref{eq_varcost}) and as summarised in Figure \ref{fig_sampling_varcost}. It is a convenient proxy for the ``difficulty'' in meeting demand: timesteps with a high variable cost are exactly those requiring the most dispatchable generation capacity.

The optimal capacities across samples of varying total size $N^S$ are determined. Each sample includes the $N^H=60$ timesteps with the highest estimated variable cost and a random sample of those remaining. For additional clarity, the steps introduced in Section \ref{doc_method} applied to the test case model are explicitly presented below.
\begin{enumerate}
\item Randomly sample $N^S$ timesteps from the 36-year dataset to create an equally weighted stage 1 subsample $(d_t, w_t, \text{weight}_t)_{t \in \mathcal{T^\text{S1}}}$
\item Determine optimal capacities:
  \begin{equation}
    [\text{cap}_b^\text{S1}, \text{cap}_m^\text{S1}, \text{cap}_p^\text{S1}, \text{cap}_w^\text{S1}] = \text{PSM} ((d_t, w_t, \text{weight}_t)_{t \in \mathcal{T^\text{S1}}}).
  \end{equation}
\item Estimate variable cost of each timestep in full dataset using equation (\ref{eq_varcost}). The variable cost is the specific choice of \textit{importance} function:
  \begin{align}
    \text{varcost}_t &= \text{varcost}(d_t, w_t,  [\text{cap}_b^\text{S1}, \text{cap}_m^\text{S1}, \text{cap}_p^\text{S1}, \text{cap}_w^\text{S1}]) \\
    &= f_b \text{gen}_{bt} + f_m \text{gen}_{mt} + f_p \text{gen}_{pt}.
  \end{align}
\item From the full dataset, construct stage 2 subsample by including the $N^H=60$ timesteps with the highest variable cost and a random sample of size $N^S - N^H$ from those remaining. Weight the timesteps in each of the 2 bins to reflect their relative proportions throughout the full dataset:
  \begin{equation}
    \text{weight}_t =
    \begin{cases}
      \frac{1}{315360} \hspace{5.3em} \text{60 timesteps with highest variable cost} \\
\frac{315360 - 60}{315360(N^S - 60)} \hspace{2em} \text{otherwise}.
     \end{cases}
   \end{equation}
   where 315360 = 8760$\cdot$36 is the total number of hourly timesteps in the 36-year dataset.
   \item Determine optimal capacities for the stage 2 subsample:
  \begin{equation}
    [\text{cap}_b^\text{S2}, \text{cap}_m^\text{S2}, \text{cap}_p^\text{S2}, \text{cap}_w^\text{S2}] = \text{PSM} ((d_t, w_t, \text{weight}_t)_{t \in \mathcal{T^\text{S2}}}).
  \end{equation}
  This is the estimate for the optimal system design.
\end{enumerate}

\subsection{Results}
\label{doc_tc_results}

\begin{figure}
\hspace{4.7em} Baseload \hspace{4em} Mid-merit \hspace{3.5em} Peaking \hspace{4.7em} Wind \vspace{1em} \\  
\centerline{(a) 480 timestep equivalent computational cost}
\includegraphics[scale=0.7, trim=0 5 0 22, clip]{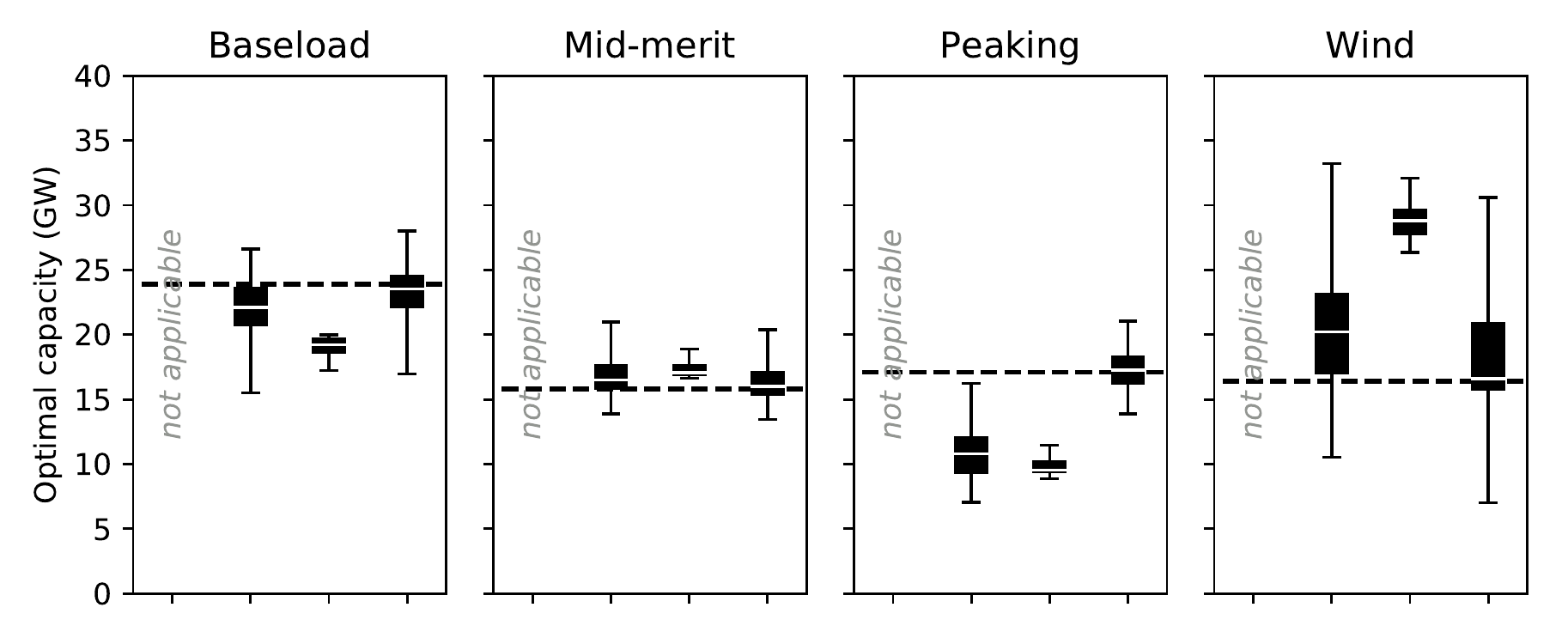} \\
\centerline{(b) 1920 timestep equivalent computational cost}
\includegraphics[scale=0.7, trim=0 5 0 22, clip]{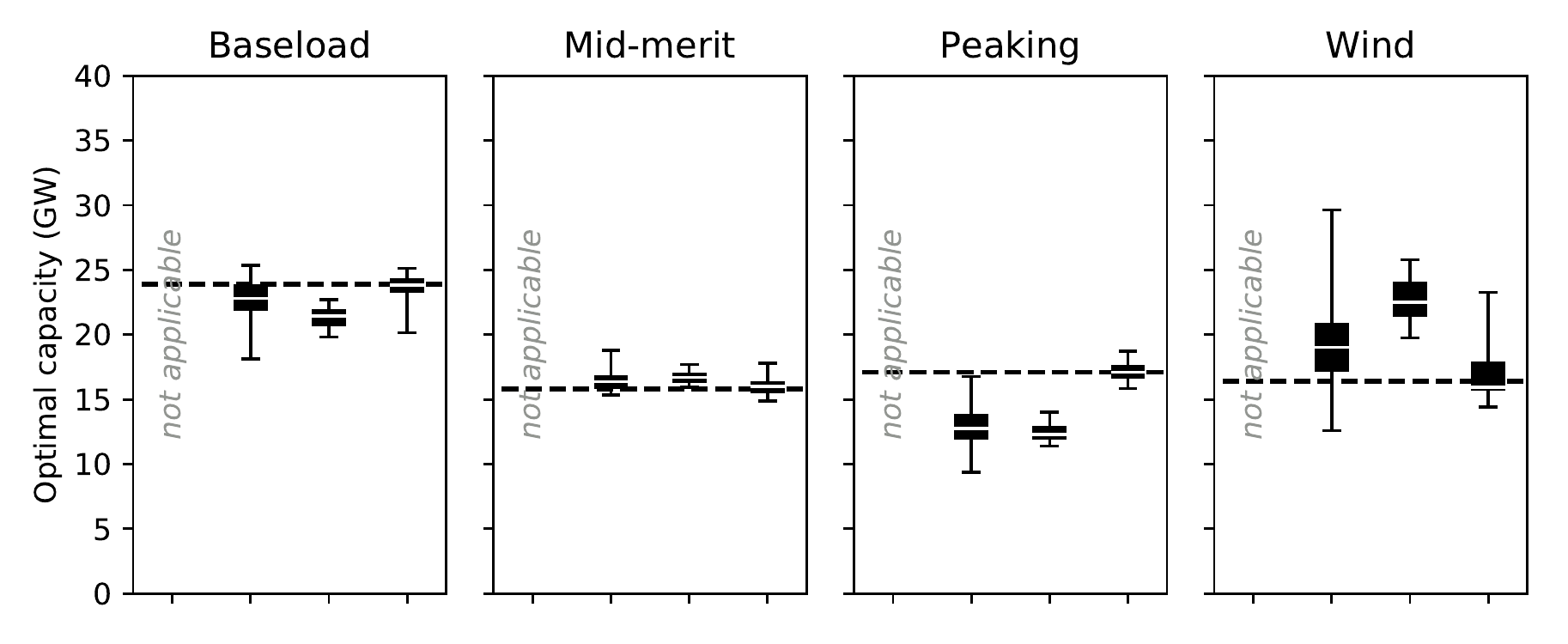} \\
\centerline{(c) 8760 timestep equivalent computational cost}
\includegraphics[scale=0.7, trim=0 10 0 22, clip]{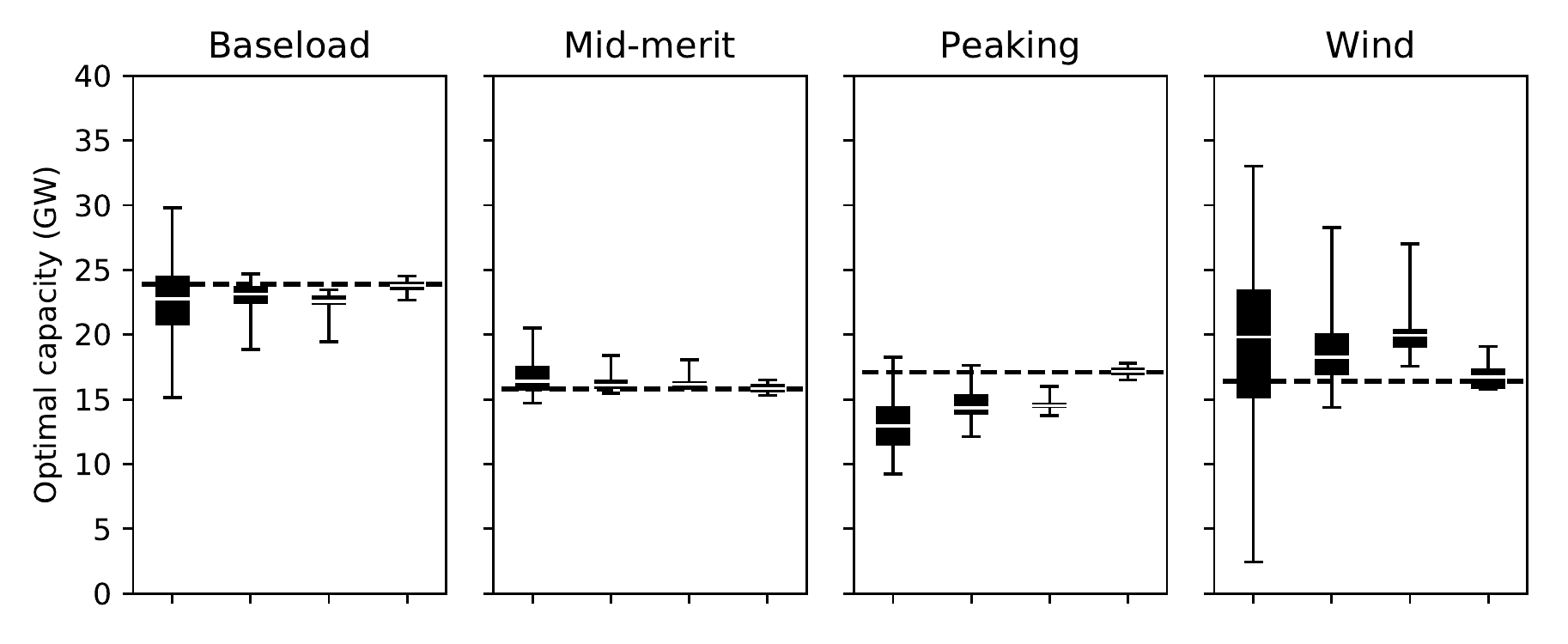} \\
\hspace*{0.2em}
\includegraphics[scale=0.7, trim=0 0 0 0, clip]{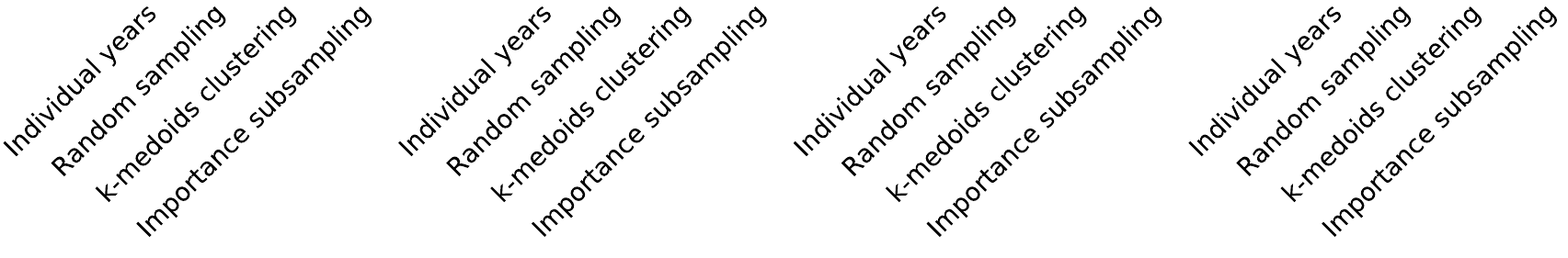}
\caption{Distribution of optimal capacities for different subsampling methodologies. The box shows the 25th, 50th (median) and 75th percentiles, while the whiskers show the 2.5th and 97.5th. (a) corresponds to a computational cost equivalent to a single PSM run using 480 timesteps, while (b) and (c) correspond to 1920 and 8760 timesteps respectively. The dashed line indicates the optimal capacities across all 36 years of data: the best estimate of the ``true'' optima and the target under subsampling.}
\label{fig:caps_comparison}
\end{figure}

\begin{figure}
\hspace{4.8em} Hours of unmet demand \hspace{8.5em} Extra system cost \vspace{1em} \\  
\centerline{(a) 480 timestep equivalent computational cost} \vspace{-0.7em} \\
\includegraphics[scale=0.7, trim=0 5 0 22, clip]{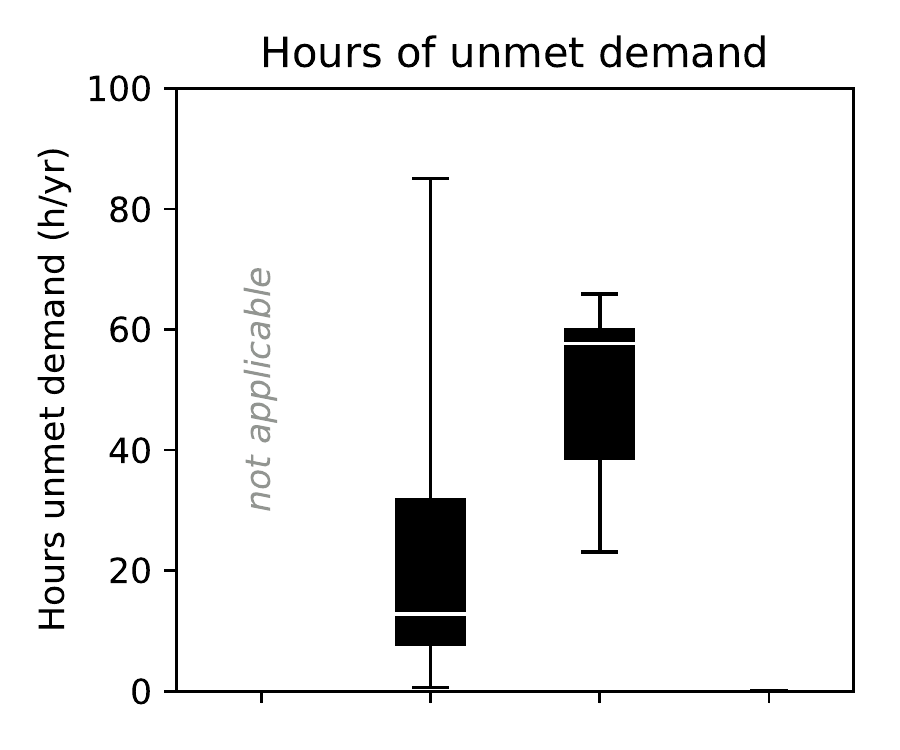}
\includegraphics[scale=0.7, trim=0 5 0 22, clip]{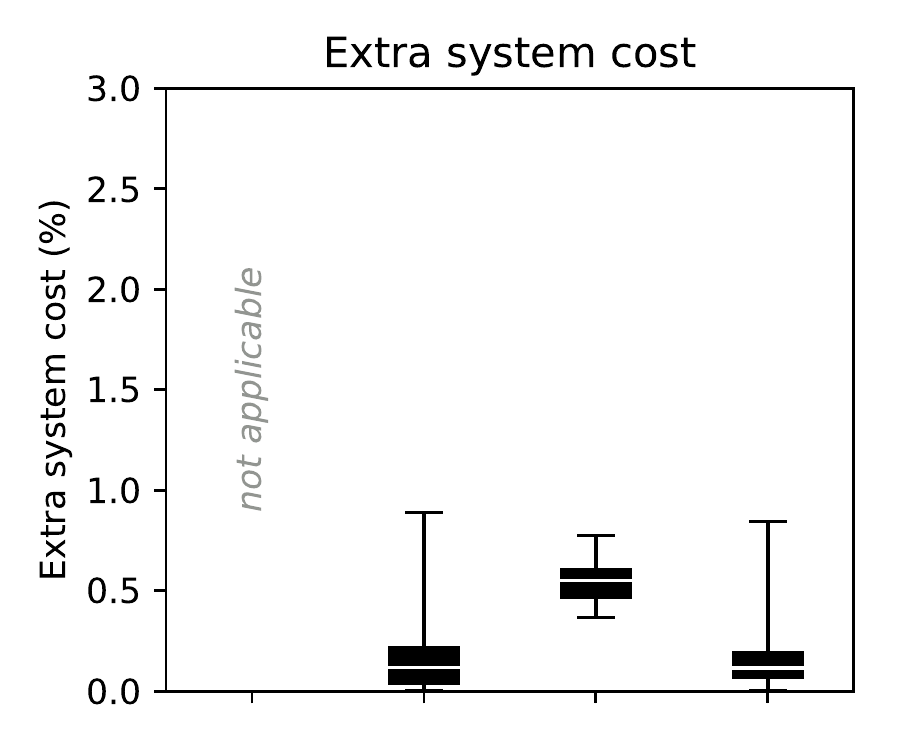} \\
\centerline{(b) 1920 timestep equivalent computational cost} \vspace{-0.7em} \\
\includegraphics[scale=0.7, trim=0 5 0 22, clip]{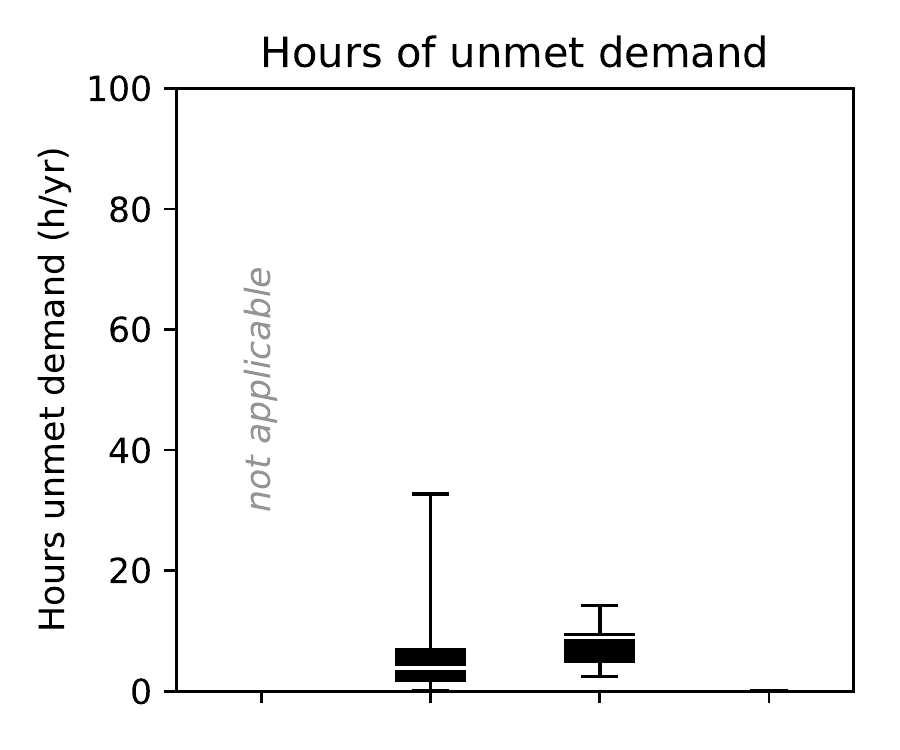}
\includegraphics[scale=0.7, trim=0 5 0 22, clip]{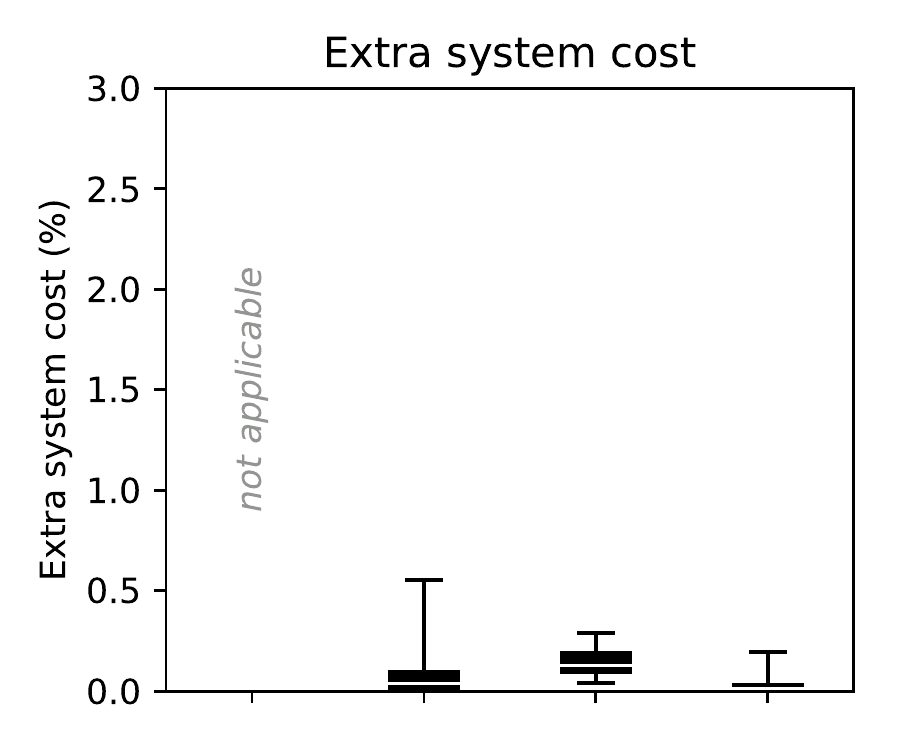} \\
\centerline{(c) 8760 timestep equivalent computational cost} \vspace{-0.7em} \\
\includegraphics[scale=0.7, trim=0 10 0 22, clip]{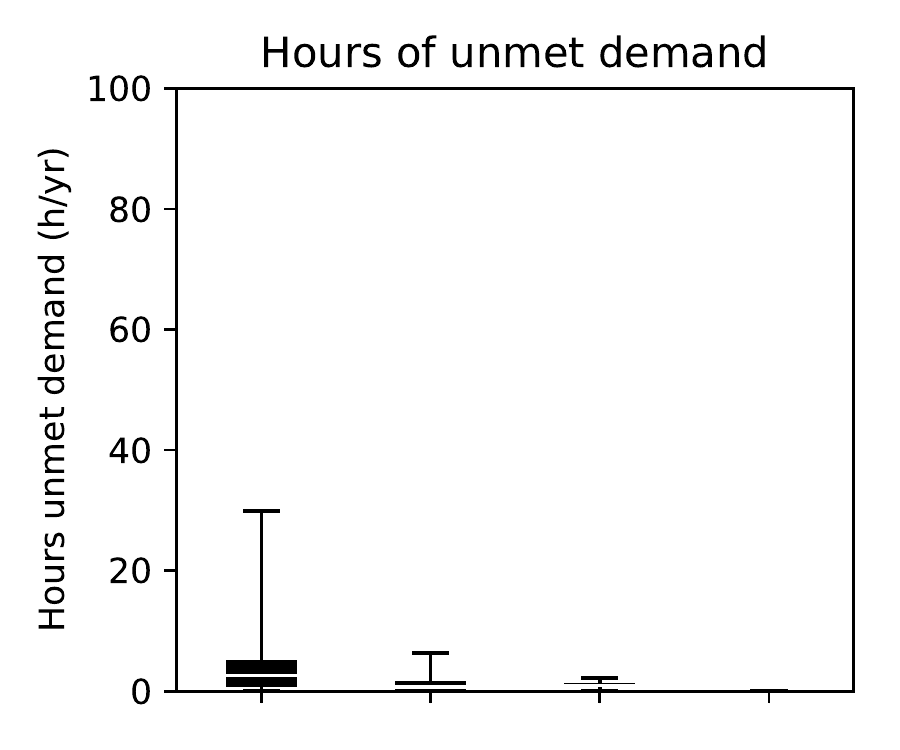}
\includegraphics[scale=0.7, trim=0 10 0 22, clip]{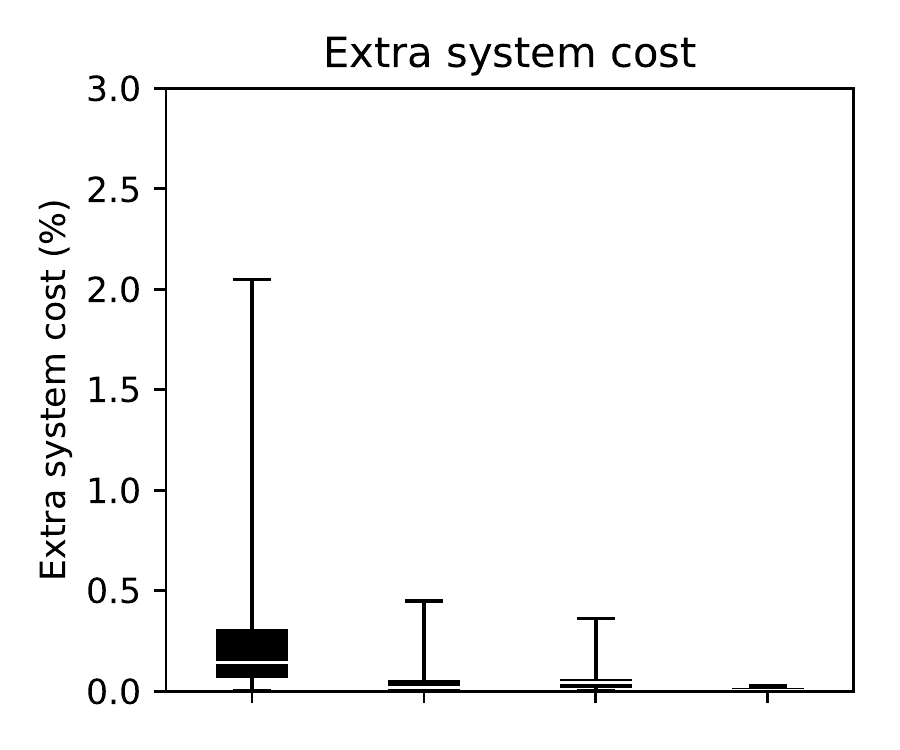} \\
\hspace*{1em} \includegraphics[scale=0.7, trim=0 0 0 0, clip]{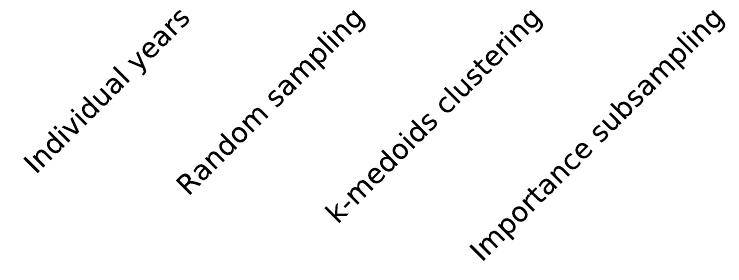} \hspace{3em}
\includegraphics[scale=0.7, trim=0 0 0 0, clip]{fig5_legend.pdf}
\caption{Distribution of \textit{hours of unmet demand} and \textit{extra system cost} for different subsampling methodologies. The box shows the 25th, 50th (median) and 75th percentiles, while the whiskers show the 2.5th and 97.5th. (a) corresponds to a computational cost equivalent to a single PSM run using 480 timesteps, while (b) and (c) correspond to 1920 and 8760 timesteps respectively and the target under subampling.}
\label{fig:suboptstats_comparison}
\end{figure}

Figure \ref{fig:caps_comparison} shows the distribution of optimal capacites across samples generated via different subsampling schemes. The 36-year optima (the targets) are shown as dashed black lines. Plots compare capacities obtained at equal computational cost. Since determining a single set of capacities using \textit{importance subsampling} requires two optimisation runs and for which the computational expense scales linearly in the number of timesteps, the associated plots use half the sample length. For example, in Figure \ref{fig:caps_comparison}(c), the box-and-whiskers plots for individual years, random sampling and $k$-medoids clustering correspond to single simulations using 8760 hourly timesteps, while the plot for \textit{importance subsampling} uses two runs of half this length each. Individual years have only one possible sample length of 8760 hourly timesteps. Since $k$-medoids clustering subsamples representative days, $k$ is the sample size divided by 24 (e.g. 480 timesteps correspond to 20 representative days).

Figure \ref{fig:caps_comparison}(a) shows the distribution of optimal capacities across simulations with computational expense equivalent to a PSM run of length 480 timesteps. While the variabilities across capacities determined via random sampling and \textit{importance subampling} are roughly equal, the former induces a larger bias (understood as the difference between the median, indicated by the white line inside the box, and the 36-year optimum, indicated by the dashed black line). $k$-medoids clustering into 20 representative days leads to low variabilities but large biases, with underestimation of optimal baseload and peaking capacity of 5 and 8GW respectively and overestimation of optimal wind capacity by more than 10GW.

Figure \ref{fig:caps_comparison}(b) shows the same plots for computational expense equivalent to 1920 timesteps (just under 3 months of hours). \textit{Importance subsampling} performs better than random sampling and clustering, which again underestimate (overestimate) optimal peaking (wind) capacity respectively. 

For simulations corresponding to full years of computational cost (8760 timesteps, Figure \ref{fig:caps_comparison}(c)), \textit{importance subsampling} performs markedly better than the other schemes. A 95\% prediction interval across individual-year simulations ranges from 2.5GW to 33.0GW, giving the user virtually no indication of optimal design. Results are slightly better under random sampling and $k$-medoids clustering. However, for all 3 schemes, the majority of samples underestimate optimal peaking capacity and overestimate optimal wind capacity. For \textit{importance subsampling}, bias is low (the median of optima lies almost exactly on the target) and variation is greatly reduced.

Figure \ref{fig:suboptstats_comparison} shows the distribution of the number of hours of annual unmet demand. Power systems designed using individual years, random sampling and $k$-medoids clustering have a high probability of leading to supply capacity shortages. For example, one designed using $k$-medoids clustering into 20 days (480 timesteps) has a 75\% change of unmet demand in at least 40 hours annually. Similarly, a power system designed using a random choice of individual year has a 35\% chance of failing to meet demand in more than 3 hours. In contrast, \textit{importance subsampling} leads to virtually no unmet demand, even for short samples lengths.

Figure \ref{fig:suboptstats_comparison} also shows the distribution of \textit{extra system cost}. Extra cost for \textit{importance subsampling} are similar to those for random sampling for 480 timesteps worth of computational expense but levels are lower than other samplers for longer simulation lengths. At computational cost equivalent to a 1920-timestep simulation, 95\% of simulations lead to cost overruns of no more than 0.2\%, and for 8760 timesteps any extra costs are negligible. 

\begin{figure}
\centering
\includegraphics[scale=0.75, trim=0 0 0 23, clip]{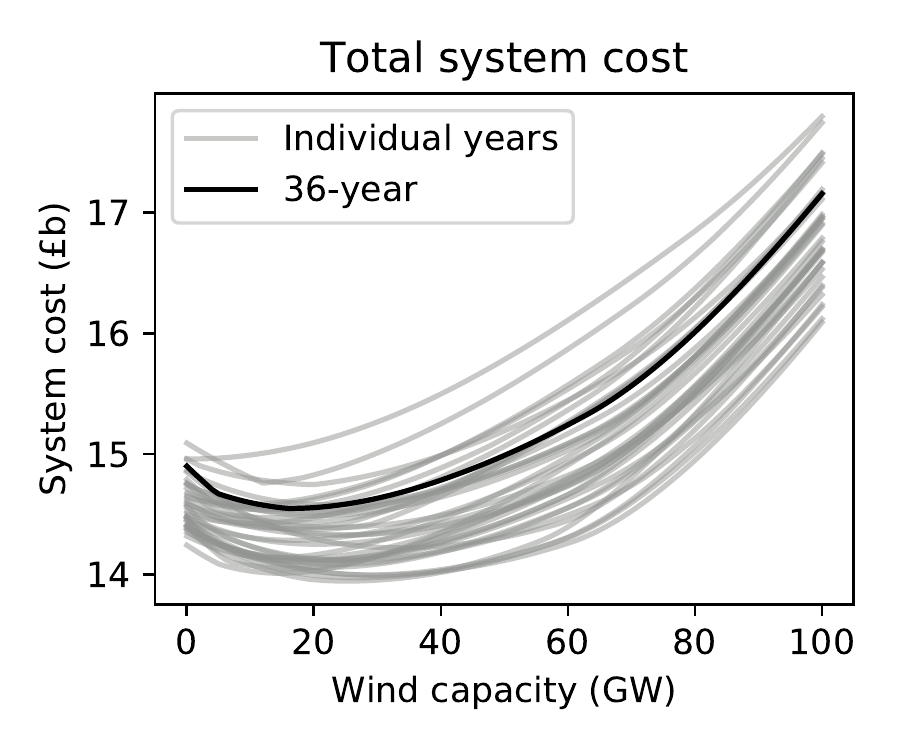} \\
\caption{Optimal system cost as a function of prescribed wind capacity, for each of the 36 years and an annualised 36-year model run.}
\label{fig:windcon_costs}
\end{figure}

The percentage errors in optimal capacities for individual technologies are much larger than those in the total system cost. For example, the 2010 optimal wind capacity is 1.0GW, a 94\% error when compared to the 36-year optimum of 16.4GW. However, the resultant power system (with extra peaking to ensure no unmet demand) costs just 2.3\% more. Across all years, the mean absolute percentage error in installed wind capacity is 34\%, but the resultant systems have a mean cost just 0.3\% higher. The reason for this is twofold. Firstly, calculating the extra system cost assumes additional peaking (to meet unmet demand) is installed at the same ``standard'' price \textemdash \, there is no additional penalty for the unanticipated need for this additional capacity. A stronger penalisation for unmet demand (such as a loss-of-load cost, typically much higher than the costs used here) would greatly increase the additional expense. Secondly, there is a flat optimum with respect to installed wind capacity, shown Figure \ref{fig:windcon_costs}. This curve is constructed by imposing a fixed wind capacity of $x$ and optimising the capacities of the remaining technologies in the usual way. The curve's $y$-value hence shows the minimum system cost as a function of prescribed wind capacity. There is one curve for each individual year along with an annualised version using all 36 years at once. In most years, choosing a wind capacity anywhere between 0 and 40GW has a small effect on cost. This explains the large inter-year variability in optimal wind capacity; changes to the annual distribution of demand and wind change the shape of the cost curves slightly and may perturb the $x$-value of the minimum significantly with a comparatively modest effect on system cost.

\section{Discussion}
\label{doc_discussion}

Previous studies indicate that, to reliably estimate optimal system design under natural climate variability, power system models (PSMs) with a high proportion of variable renewable energy require both a high temporal resolution and long samples of demand \& weather data. These combined requirements typically exceed computational limitations, and timeseries reduction approaches such as subsampling are required. The accuracy of PSM outputs using subsampled data is highly dependent on whether a number of ``extreme'' timesteps are included in modelling samples. However, identifying them is difficult \textit{a priori} since which timesteps are ``extreme'' frequently themselves depends on what the model is trying to determine (a chicken-and-egg situation).

This paper introduces a novel subsampling approach, called \textit{importance subsampling}, that systematically identifies such timesteps and includes them in modelling samples. The main idea is to use a first stage simulation to obtain a rough indication of optimal system design. This is used, in conjunction with an \textit{importance} function, to determine which timesteps must be sampled. A second stage model run, including these timesteps, provides estimates for the model outputs found using all available data.

A test case is performed on a model of the United Kingdom and 36 years of historic demand \& weather data. Optimal system design for individual years is found to be highly dependent on the choice of year as well as underestimating (overestimating) optimal dispatchable (renewable) capacity respectively. Furthermore, resultant power systems lead to supply capacity shortages when applied to different weather years, and the long-term optimal power system design cannot be reliably determined by taking averages across multiple individual-year simulations.

Estimates of optimal system design using random samples or data clustered into representative days have lower variability and slightly smaller errors. Interestingly, randomly sampling 8760 hourly timesteps more accurately estimates optimal capacities than individual years. Hence, in models where timesteps may be scrambled (e.g. ones ignoring ramping or storage constraints), random sampling may be preferred over selecting long contiguous periods. However, errors in estimation of optimal system design and associated generation capacity shortages or cost overruns are present across both random sampling and clustering. 

In contrast, \textit{importance subsampling} consistently estimates, at reduced computational cost, the model outputs found using all available data. Even for short sample lengths, the bias between the median of capacities found using \textit{importance subsamples} and the long-run optima is very small. This means a user can estimate optimal capacites by the median across multiple PSM runs with short \textit{importance subsamples}, something that would be impossible with the other subsampling approaches due to their biases.

The \textit{importance subsampling} approach is introduced in full generality and can be applied to a wide class of optimisation-based PSMs by modifying the choice of \textit{importance} function. In the test case, this is a timestep's variable cost, and this choice generalises naturally to any model with such a notion. For example, in more detailed models, the variable cost can be summed across regions or calculated considering more sophisticated techno-economic factors. A judicious choice of \textit{importance} function may also be inspired by expert knowledge. For example, if the user has a good idea of which demand \& weather scenarios are essential to ensure generation capacity adequacy, the \textit{importance} function may be chosen to identify such events.

A drawback of the proposed approach is that it scrambles the order of timesteps and hence can only be directly applied to models with a small level of continuous dependence such as storage or ramping. Using the same approach to sample longer time periods such as days or weeks allows the modelling of such phenomena within periods, which may be sufficient in many settings. This may require additional refinements such as combining \textit{importance subsampling} with clustering to reduce output variability.

The extra cost from designing suboptimal systems (and adjusting to ensure no unmet demand, see Section \ref{doc_tc_results}) is typically much smaller than the errors in estimates of optimal generation capacities. One of the causes, a flat optimal cost (particularly with respect to installed wind capacity), highlights an important distinction. If the only goal is the cheapest power system, a single-year simulation may suffice provided some extra capacity is installed to ensure demand can always be met. However, if one seeks to use minimisation of total system cost as a means of determining optimal system design, the impact of climate uncertainty is significant and approaches incorporating many years of climate information are required. In addition, one should be careful when invoking arguments of the type ``wind should not be installed since the model indicates optimal design includes almost no wind''. A more robust argument requires the investigation of the degree of suboptimality of power systems with perturbed design.

There are a number of possible extensions to this investigation. One is to relax the ``perfect model'' assumption that the 36-year simulation determines the ``true'' optimal capacities. In reality, the 36 years are themselves a sample from the ``true'' climate and the associated model outputs can be uncertain or biased. Using techinques from extreme value statistics or a climate model to generate a larger sample of hypothetical demand \& weather data offer a more robust approach. Another extension is to conduct the same investigation using a different PSM or geographical region. On the one hand, \citet{woollings_2010} notes that United Kingdom's climate exhibits a large degree of inter-year variability (so that sampling from multiple decades may not be necessary for other locations). On the other, the PSM employed in this investigation aggregates demand and wind levels into a single value for the whole country, ``averaging out'' spatial variations (and hence also some of the inter-year variability). For this reason, the degree of inter-year variability and both the need for and efficacy of \textit{importance subsampling} for other PSMs should be explored. Finally, one could investigate its use in sampling longer representative periods (e.g. days or weeks). This may require additional adjustments such as combining the method with clustering approaches to further reduce variance in model outputs.

In line with the \textit{open energy modelling} movement \citep[see][]{hilpert_2017, pfenninger_2017_2} the data and all files required for the construction of the model (in the framework \textit{Calliope}) are publicly available at \url{https://github.com/ahilbers/2019_importance_subsampling}.

\section{Appendix}
\label{doc_app}
\subsection{Power system model}
\label{doc_app_PSM}
\begin{table}
\centering
\begin{tabular}{ l p{4.5cm}}
\multicolumn{2}{l}{Sets} \\ \hline
$\mathcal{I}$ & Technologies: baseload, mid-merit, peaking, wind \\
$\mathcal{T}$ & Timesteps (hourly) \\ \hline
& \\
\multicolumn{2}{l}{Parameters} \\ \hline
$c_i$ & Annualised installation cost, technology $i$ (\pounds m/GWyr) \\
$f_i$ & Generation cost, technology $i$ (\pounds m/GWh) \\ \hline
\end{tabular}
\hspace{1em}
\begin{tabular}{ l p{4.5cm}}
\multicolumn{2}{l}{Timeseries data, timestep $t$} \\ \hline
$d_t$ & Demand (GWh) \\
$w_t$ & Wind capacity factor \\
$\lambda_t$ & Timestep weight \\ \hline
& \\
\multicolumn{2}{l}{Decision variables} \\ \hline
$\text{cap}_i$ & Installed capacity, technology $i$ (GW) \\
$\text{gen}_{it}$ & Electricity generated, technology $i$, timestep $t$ (GWh) \\ \hline
\end{tabular}
\caption{Nomenclature used to describe the PSM in Section \ref{doc_app}.}
\label{table_ra_nomenclature}
\end{table}

The PSM is a simple representation of the UK power system, viewing it as a single node with inelastic UK-wide hourly demand levels and wind capacity factors. A linear optimiser determines the optimal capacities across 4 possible technologies by minimising the sum of installation and generation costs. The first 3, generically called \textit{baseload}, \textit{mid-merit} and \textit{peaking}, differ only in their fixed (investment, \pounds/GW) and variable (generation, \pounds/GWh) costs. The fourth option, \textit{wind}, has 0 generation cost but a supply level capped by the installed capacity times the wind capacity factor. Its installation cost refers to \textit{rated capacity}: the maximum power output that a wind farm can produce at wind speeds just under the \textit{cut-out point} when turbines are turned off to avoid damage. The model takes two timeseries inputs: hourly UK-wide demand levels and wind capacity factors.

\subsubsection{Assumed costs}
\begin{table}
\centering
\begin{tabular}{ l  c  c}
& Installation cost & Generation cost \\
Technology &(\pounds m/GWyr) &(\pounds m/GWh) \\ \hline
Baseload & $c_b = 300$ & $f_b = 0.005$ \\
Mid-merit & $c_m = 100$ & $f_m = 0.035$ \\
Peaking & $c_p = 50$ & $f_p = 0.1$ \\
Wind & $c_w = 100$ & $f_w = 0$ \\ \hline
\end{tabular}
\caption{Assumed installation and generation costs for the 4 technologies used in the power system model employed in this paper.}
\label{table_ra_costs}
\end{table}

Table \ref{table_ra_costs} shows each technology's assumed costs. They are chosen to be roughly indicative of the possible generation technologies available in the UK but should not be interprested as realistic estimates of future build or generation cost. Each technology's installation cost is annualised by dividing by the expected plant lifetime.

Mid-merit prices are based on the CCGT H class reactor from \citep{beis_2016}, which has a build cost of \pounds100/kW per year of plant lifetime, corresponding to $c_m=$ \pounds100m/GWyr. The generation cost is determined assuming a wholesale gas price of 60 pence per therm. Using the conversion 1 therm $\approx$ 30kWh and assuming a generation efficiency of 55\% gives a generation cost $f_m \approx$ \pounds 35t/GWh. The costs for baseload and peaking are chosen either side of these values.

Wind is assumed to have no generation cost. The construction cost is taken from the \textit{medium} value for an \textit{Onshore UK $>$5MW} wind farm in \citep{beis_2016}: \pounds1200/kW, corresponding to \pounds50/kWyr over a 24-year lifetime. Adding fixed O\&M costs of \pounds23k/kWyr gives \pounds73m/GWyr. This is revised upwards to \pounds100m/GWyr to reflect infrastructure costs and the fact that offshore wind is more expensive than onshore.

\subsubsection{Mathematical setup}
The optimiser chooses capacities by minimising the sum of installation and generation costs while meeting an inelastic demand timeseries, phrased as a linear optimisation problem with decision variables $\{\text{cap}_i, \text{gen}_{it}, \, i \in \mathcal{I}, t \in \mathcal{T}\}$. Timesteps are 1 hour in length.
\begin{equation}
\min_{\{\text{cap}_i, \text{gen}_{it}, \, i \in \mathcal{I}, \, t \in \mathcal{T} \}} z = \sum_{i \in \mathcal{I}} \left[ c_i \text{cap}_i + 8760 \sum_{t=1}^{T} f_i \lambda_i \text{gen}_{it} \right]
\label{3techsw_objective}
\end{equation}
\noindent subject to 
\begin{align}
\text{gen}_{it} \le \text{cap}_i \quad & \forall \: i \in \mathcal{I} \backslash \{w\}, \: \forall \: t \in \mathcal{T}
\label{3techsw_generation_less_capacity_conv} \\
\text{gen}_{wt} \le \text{cap}_w w_t \quad \: & \forall \: t \in \mathcal{T}
\label{3techsw_generation_less_capacity_wind} \\
\sum_{i \in \mathcal{I}} \text{gen}_{it} = d_t \quad & \forall \: t \in \mathcal{T}
\label{3techsw_demand_meets_supply} \\
\text{cap}_i, \text{gen}_{it} \ge 0 \quad & \forall \: i \in \mathcal{I}, \: \forall \: t \in \mathcal{T}.
\label{3techsw_ge_0}
\end{align}

\noindent For definitions of letters and symbols, see Table \ref{table_ra_nomenclature}. The constant factor 8760 (= the number of hours in a year) is included since weights are chosen to sum to 1.  

Constraint (\ref{3techsw_generation_less_capacity_conv}) ensures that conventional (baseload, mid-merit and peaking) generation levels never exceed their capacity.  Constraint (\ref{3techsw_generation_less_capacity_wind}) ensures that wind generation never exceeds installed capacity times the wind capacity factor. Constraint (\ref{3techsw_demand_meets_supply}) ensures that supply meets demand in each timestep. Given installed capacities, generation levels are determined by merit-order stacking of technologies in ascending order of variable cost.

The PSM is built and solved in the open-source modelling framework \textit{Calliope} \citep[see][]{pfenninger_2018}.

\subsection{Data}
\label{doc_app_data}
This investigation uses two timeseries: hourly UK-wide electricity demand levels $(d_t)_{t \in \mathcal{T}}$ and wind capacity factors $(w_t)_{t \in \mathcal{T}}$ over the period 1980-2015. They are modified versions of those found in \citep{bloomfield_2016}

\subsubsection{Demand model}
\label{doc_app_data_demand}
The demand timeseries is based on a regression between weather and demand data collected from two different sources. UK-wide daily mean temperature is obtained from the MERRA reanalysis \citep{merra_2017} for the period 1980-2015. Metered UK-wide demand data is obtained from National Grid, the UK transmission operator, over 2006-2015 \citep[see][]{national_grid_data}. For this period, in which there is overlapping meteorological and demand data, a regression model is run for daily demand:
\begin{multline}
\textrm{Demand}(t_d) = \alpha_1 + \alpha_2t_d + \alpha_3 \textrm{sin}(\omega t_d) + \alpha_4 \textrm{cos}(\omega t_d) + \alpha_5 \textrm{Te}(t_d) + \alpha_6 \textrm{Te}^2(t_d) \\
+ \sum_{k=7}^{13} \alpha_k \textrm{WD}_k(t_d) + \alpha_{14} \textrm{HOL}(t_d) + \epsilon(t_d)
\label{eq_demandmodel}
\end{multline}
where 
\begin{itemize}
\item $\text{Demand}(t_d)$ is the totaly daily demand in day $t_d$.
\item $\alpha_1 + \alpha_2t_d$ accounts for anthropogenic demand trends (e.g.\ economic growth or efficiency improvements).
\item $\omega$ is $\textrm{year}^{-1}$.
\item $\sin(\omega t_d)$ and $\cos(\omega t_d)$ account for year-long demand cycles.
\item Te($t_d$) is the \textit{effective temperature}: a slightly time-offset daily mean temperature to account for the fact that demand is influenced by temperature at a time lag. It is defined recursively via $\text{Te}(t_d) = \frac{1}{2}\text{Te}(t_d - 1) + \frac{1}{2}\text{T}(t_d)$, where $T(t_d)$ is the UK-wide daily average temperature on day $t_d$, and initialised via $\text{Te}(t_d = \text{1-1-2006}) = \text{T}(t_d = \text{1-1-2006})$.
\item $\textrm{WD}_7(t_d)$ and $\textrm{WD}_{13}(t_d)$ are indicator variables that take value 1 only on Monday through Sunday respectively, but not on bank holidays.
\item $\textrm{HOL}(t_d)$ is an indicator variable that takes value 1 only on bank holidays.
\item $\epsilon (t_d)$ is an error term.
\end{itemize}

For values of the parameters, see \citep{bloomfield_2016}. The distribution of the error term $\epsilon(t_d)$ is symmetrical with a sharp mode at 0 and a standard deviation of 34.2GWh. There is no trend in average daily error across years or months.

After determining the coefficients, daily demand is extrapolated to 1980-2015 by truncating $\epsilon(t_d)$ and using the daily temperatures to extend the timeseries back. Four diurnal curves, one for each of the 3-month periods DJF, MAM, JJA, SON, are used to upsample the timeseries to hourly resolution, with each day given a weighting of two diurnal curves. For example, December 1 is given a 50/50 weighting of the diurnal curves for SON and DJF.

Anthropogenic demand trends are removed by setting $\alpha_2$=0 and replacing $\alpha_1$ by $\hat{\alpha}_1 = \alpha_1 + \alpha_2 t | _{t=1-1-2011}$, representing the middle of the period 2006-2015. This ensures that the differences between the demand distribution in different years is weather-driven.

\subsubsection{Wind model}
\label{doc_app_data_wind}
The wind model is identical to that used in the \textit{HIGH wind} scenario in \citep{bloomfield_2016}, which is in turn based on two other studies. First, from the MERRA re-analysis \citep{merra_2017}, hourly windspeeds at 2, 10 and 50m elevation are obtained at gridpoints throughout the UK. These are fed through a logarithmic profile to estimate windspeeds at 80m, the average height of UK wind turbines. 

Windspeeds at MERRA gridpoints are linearly interpolated to the locations of all wind farms in operation or in the construction pipeline as of April 2014 \citep[see][]{drew_2015}. At each location, wind speeds are transformed to capacity factors using an adjusted power curve from a Siemens 2.3MW turbine \citep[see][for details]{cannon_2015}. An average over the wind farm configuration, weighted by each farm's capacity, determines the UK-wide capacity factor. In the PSM, adding capacity happens in proportion to the wind farms' relative sizes. For example, doubling wind capacity implicitly means doubling the capacity of each wind farm.

\subsection{Supplementary material}
\label{doc_app_supp}

The demand timeseries used in the test case (Section \ref{doc_tc}) as introduced in Section \ref{doc_app_data_demand} is based on a regression. In particular, in obtaining 36 years of demand data, the error term in equation (\ref{eq_demandmodel}) is truncated. This has the effect of reducing variability. For this reason, the performance of \textit{importance subsampling} is also examined on a real demand timeseries, where only a long-term demand trend is removed but there is no truncation of the error. Figures \ref{fig:caps_comparison} and \ref{fig:suboptstats_comparison} are recreated using the altered demand timeseries. Results are broadly identical as those found using regression data and are available as supplementary material at \url{github.com/ahilbers/2019_importance_subsampling}.

\section*{Acknowledgements}
The first author's work was funded through the EPSRC CDT in Mathematics for Planet Earth. The authors thank Hannah Bloomfield for providing demand and wind timeseries and 3 anonymous referees for their constructive comments.

\section*{References}
\bibliography{citations}

\end{document}